\pdfoutput=1
\documentclass[aps,pra,10pt,twocolumn,showpacs,superscriptaddress,floatfix]{revtex4-1}
\usepackage[usenames,dvipsnames,svgnames]{pstricks}

\usepackage{newlfont}
\usepackage{amssymb}
\usepackage{amsfonts}
\usepackage{amsmath}
\usepackage{amsthm}
\usepackage{graphicx}
\usepackage{color}
\usepackage{bm}
\usepackage{times}
\usepackage[draft]{hyperref}

\begin{document}
\title{Non-commutative space engine: a boost to thermodynamic processes}

\author{Tanmoy Pandit}
\email{tanmoypandit163@gmail.com}
\affiliation{Hebrew University of Jerusalem, Jerusalem 9190401, Israel}

\author{Pritam Chattopadhyay}
\email{pritam.cphys@gmail.com}
\affiliation{Cryptology and Security Research Unit, R.C. Bose Center for Cryptology and Security,\\
Indian Statistical Institute, Kolkata 700108, India}

\author{Goutam Paul}
\email{goutam.paul@isical.ac.in}
\affiliation{Cryptology and Security Research Unit, R.C. Bose Center for Cryptology and Security,\\
Indian Statistical Institute, Kolkata 700108, India}


\begin{abstract}
We introduce quantum heat engines that perform quantum Otto cycle and the quantum Stirling cycle by using a coupled pair of harmonic oscillator as its working substance. In the quantum regime,  different working medium is considered for the analysis of the engine models to boost the efficiency of the cycles. In this work, we present Otto and Stirling cycle in the quantum realm where the phase space is non-commutative in nature. By using the notion of quantum thermodynamics we develop the thermodynamic variables in non-commutative phase space. We encounter a catalytic effect (boost) on the efficiency of the engine in non-commutative space (i.e, we encounter that the Stirling cycle reaches near to the efficiency of the ideal cycle) when compared with the commutative space. Moreover, we obtained a notion that the working medium is much more effective for the analysis of the Stirling cycle than that of the Otto cycle.
\end{abstract}

\maketitle

\section{Introduction}\label{sec1}
Non-commutativity is one of the possible theories for the general structure of the spacetime. One can describe the quantization of manifold by utilizing the language of non-commutative (NC) geometry. The seminal work of Gelfand-Naimark \cite{gel} has shown a one-to-one relation between commutative algebra and certain space structures. The idea behind the non-commutative geometry is to inculcate non-commutative algebras as non-commutative geometric spaces. To ameliorate the ultraviolet singularity in Quantum Filed Theory, Snyder~\cite{hss} pioneered the idea of a short distance cutoff scale using NC spacetime. This work obtained its popularity when Seiberg and Witten~\cite{sei} in their work utilized NC spacetime for the analysis of the open strings with endpoints fixed on ``$D-$brans" at low energy limit. To analyze and predict quantum gravity in low energy phenomena~\cite{das,ali,tka,ghoshs}, further forms of NC spacetime were formulated. The formalism of NC spacetime predicts the existence of the minimum length. It gives rise to a contradictory statement with the Heisenberg uncertainty principle~\cite{heisn}. Heisenberg's uncertainty principle conveys that the uncertainty in position can be infinitely small. In NC spacetime, Generalized Uncertainty Relation (GUP)~\cite{das,kemp,souvik,souvik1,husain,todo} furnishes the description of the existence of the minimum length. The minimum length is coined as Planck length.

In theoretical physics, thermodynamics has a strong ground of its own. In recent times, exploration of thermodynamics in the quantum regime, namely, quantum thermodynamics~\cite{gemm} have acquired its importance. A great effort has been devoted to the exploration and discussions about the derivation of the second law of thermodynamics in the quantum regime or deriving the whole theory of thermodynamics from the quantum information theory viewpoint~\cite{goo}. In previous works, the definition of the second law of thermodynamics in the presence of an ancilla~\cite{horo,fbr}, or system which have coherence~\cite{mlo,mlo1} has been explored in great details. Under appropriate limits,  one can produce the classical version of the second law from the quantum version. 

In quantum realm, one can take different approaches for the analysis of thermodynamic processes such as information-theoretic viewpoint~\cite{rla,chb,elu,kma,ldr,gt} or resource-theoretic approach~\cite{fgs,mhor,ggo}.  One of the prime constituent areas of study in this direction is the work extraction from quantum systems~\cite{jab,psk,mpe}. An intense focus on quantum thermodynamic machines provides new insight. It was during the 1950's when the first work~\cite{hed} in this direction appeared but it regained its interest when Kieu~\cite{tdk,tdk1} in his work, proposed a quantum heat engine by using a single two-level system as a working medium. 

The study of quantum thermodynamic machines assists us to interpret the behavior of the thermodynamic quantities in the quantum realm like work, efficiency, heat due to the non-classical features that occur in the quantum regime such as entanglement, quantum superposition, squeezing~\cite{mos,rko,jro1}. This type of thermodynamic machine has practical importance in the field of quantum computation and refrigeration in micro regimes~\cite{nli}. Coupled quantum systems, as the working medium for heat engines, have been studied widely in previous works~\cite{gtho,thom}. It is shown in the work~\cite{gtho,thom}, that with appropriate coupling the efficiency of the system can be increased compared with the uncoupled one. Experimental realization of Otto cycle~\cite{b22} has also been analyzed.

In this work, we have considered a coupled harmonic oscillator as the working substance for the analysis of quantum cycles in commutative space and non-commutative \textcolor{black}{space}. We have considered two non-commutative phase space structure for our analysis. Our ultimate goal is to inspect different quantum engine cycles with coupled harmonic oscillators as the working substance for considered non-commutative phase-space structures and also for the commutative space structure. Different heat cycles in the quantum realm have  harmonic oscillator as its working principle. We have considered two reversible cycles, i.e., the Stirling cycle and the Otto cycle for analysis. \textcolor{black}{The Otto cycle is a paradigmatic choice for analysis, but in recent times the study of the quantum Stirling cycle has gained importance. In various works~\cite{rv13,rv14,rv15}, analysis of both the engine model on the same ground, i.e., with the same physical systems and environment are performed. In the work~\cite{rv13}, they have shown that we can expect better efficiency for the Stirling cycle than that of the Otto cycle, and it has been recently shown in the work~\cite{rv15} that small perturbation in the harmonic oscillator can enhance the performance of quantum refrigerators for both the Otto and Stirling cycles. So, having these preliminary findings, we have considered both the engine models for our analysis to verify whether the Stirling cycle has an advantage over the Otto cycle when the working systems are in the non-commutative space.} The efficiency is evaluated for each engine cycle in commutative and NC phase space structure after the working substance evolves through every stage of their individual cycles. The effects are astonishing when the cycles are in NC phase space. 

In the case of the Otto cycle, the coupling strength of the coupled oscillator produce a constant efficiency in commutative phase space but it gets a boost when the engine is in NC phase space. Similarly, when the Stirling cycle is analyzed with the coupled harmonic oscillator as the working medium the coupling strength results in higher efficiency than the decoupled oscillator. But in NC phase space the efficiency gets a boost and picks up the efficiency near to the ideal models of the engine cycles. The working medium is much more effective for the Stirling cycle than the Otto cycle in all forms of space structure that is analyzed in this work. Works with space structure with different approach is shown in~\cite{san1,san2,san3,chatt123}. 

Though it seems to be mathematically feasible, one immediate question that comes to one's mind is regarding the physical accessibility of the NC phase space with the so-far existing modern quantum technology. Recent works for the physical accessibility of the NC phase space using quantum optics~\cite{ipik,sdey} and Opto-mechanical~\cite{mkho} setup has been developed. So, the possibility of experimental verification and analysis of NC phase space will provide a boost for the exploration of quantum information theory in NC space structure. 

 The prime focus of the current work is to analyze how the change in the space structure can affect the efficiency of the different quantum thermal engine models. For this work, we have considered the traditional formalism of thermodynamics. The considered system is analyzed when the system reaches its equilibrium state. The non-equilibrium thermodynamics of the engines which is referred to as finite time thermodynamics has not been explored here. Non-equilibrium thermodynamics in NC space has been explored in some previous works~\cite{girotti,dias1} that are mainly focused on the analysis of the Brownian motion in NC space structure. They have shown that the master equation for the NC space boils down to  the master equation of the ordinary space (commutative space) when the non-commutative  parameter is equated to zero.  Due to the non-commutative parameter we encounter some extra terms which affects the results that we get from the ordinary space. The non-equilibrium thermodynamics of the engine models in NC space is an open area for exploration. \textcolor{black}{Along with that for our analysis, we have considered the dimension of the bath big enough so that the system has a continuous spectrum. So, we can consider that the relaxation time of the bath is quite small than that of the system. From this condition, we can usually take into account that the system gets easily thermalized to the temperature of the bath. }

The paper is categorized in this manner: in section~\ref{sec2}, we have analyzed a coupled harmonic oscillator. Firstly it is analyzed in commutative space and then its analysis is extended to the NC space for two different models of NC space.  Section~\ref{sec3} is dedicated to the establishment of the quantum Otto cycle in commutative, and NC space with coupled harmonic oscillator as the working medium. Section~\ref{sec4} is devoted to the discussion of the Stirling cycle and its analysis in commutative space and NC space with coupled harmonic oscillator as the working medium. Here in the section~\ref{sec3} and ~\ref{sec4}, we generate the efficiency of the cycle for the different space structure models and analyze the effect of this space on the efficiency of the engines considered for the analysis. We conclude our paper in section~\ref{sec5} with some discussion.

\section{Coupled Harmonic Oscillator}\label{sec2}
We will examine a coupled harmonic oscillator (HO)~\cite{jellal,linb} system specified by the coordinates $x_1, x_2$ and masses $m_1, m_2$. One can describe this using the Hamiltonian as the sum of free and interacting parts

\begin{equation}\label{a1}
H_\alpha = {p^{2}_{1}\over 2m_{1}} + {p^{2}_{2}\over 2 m_{2}} +
{1\over 2} \left( C_1x^{2}_{1} + C_2 x^{2}_{2} + C_3 x_{1}
x_{2}\right),
\end{equation}
where $C_1, C_2, C_3$ are constant parameters and $p_1, p_2$ are the momentum of the two oscillators.
Re-scaling the position variables of the oscillators 
\begin{eqnarray}\label{a2} 
x_\alpha = \Big({m_{1}\over m_{2}} \Big)^{1\over 4} x_{1}, \quad
x_ \varpi= \Big({m_{2} \over m_{1}} \Big)^{1\over 4} x_{2},
\end{eqnarray}
and similarly for the momentum we have
\begin{eqnarray}\label{a3} 
p_\alpha = \Big({m_{2}\over m_{1}} \Big)^{1\over 4} P_{1},\quad
p_ \varpi = \Big({m_{1} \over m_{2}} \Big)^{1\over 4} P_{2}.
\end{eqnarray}

So, the Hamiltonian $H_\alpha$ in Eq.~\eqref{a1} using Eq.~\eqref{a2}, ~\eqref{a3} takes the form 
\begin{equation}\label{a4}
H_{\alpha_1} = {1\over 2m} (p^{2}_\alpha + p^{2}_ \varpi) + {1\over
2} ( c_1 x_\alpha^{2} + c_2 x^{2}_ \varpi + c_3 x_\alpha x_ \varpi),
\end{equation}
where the parameters take the form 
\begin{eqnarray}\label{a5}\nonumber
m  =  (m_{1}m_{2})^{1/2},\quad c_3 =  C_3, \\ \nonumber
\quad c_1  =  C_1\sqrt{m_2\over m_1}, \quad c_2  =  C_2\sqrt{m_1\over m_2}.
\end{eqnarray}

The Hamiltonian~\eqref{a4} represents the interaction between the two oscillators. The analysis of the system for this Hamiltonian is not so straightforward. To streamline the situation we transform to new phase variables 
\begin{equation}\label{a6} 
y_i = M_{ij} x_j, \qquad  q_i = M_{ij} p_j,
\end{equation}
where $M_{ij}$ takes the form 
$M_{ij} = \begin{pmatrix}
cos {\theta\over 2}  && -\sin {\theta\over 2} \\
\sin {\theta\over 2} && \cos {\theta\over 2}
\end{pmatrix}$. Here, $M_{ij}$ is a unitary rotation operator with the angle $\theta$. Using this transformation~\eqref{a6} the Hamiltonian~\eqref{a4} takes the form 
\begin{equation}\label{a7}
 H_{final} = {1\over 2m} (q^{2}_\alpha + q^{2}_ \varpi) +
{K\over 2} (e^{2\zeta } y^{2}_{\alpha} + e^{-2\zeta }
y^{2}_{ \varpi}),
\end{equation}
where  $K = \sqrt{c_1c_2 - c_3^{2}/4}, \,\, e^{\zeta}= \frac {c_1 + c_2 + \sqrt{(c_1 - c_2)^{2} + c_3^{2}}}{2K}$. Here $e^{\zeta}$ describes the coupling between the two coupled oscillators. The Hamiltonian has to satisfy the conditions $4c_1c_2 > c_3^{2}$ and $\alpha  = {c_3\over c_2 - c_1}$. Solving the Hamiltonian~\eqref{a7} for the eigenvalues we get 
\begin{equation}\label{a8}
 E_{n_1,n_2} = {\hbar \omega_\vartheta} \left(e^{\zeta} \left(n_1
+{1\over 2}\right) +e^{-\zeta} \left(n_2 +{1\over 2}\right)\right).
\end{equation}
Eq.~\eqref{a8} represents the energy spectrum of the two coupled harmonic oscillators in commutative space and $\omega_\vartheta$ represents the frequency of the oscillator.

%%%%%%%%%%%%%%%%%%%%%%%%%%%%%%%%%%%%%%%%%%%%%
\subsection{Coupled HO for non-commutative space}\label{Sec2a}
Now, we will analyze two coupled HO in NC spacetime. Based on the Heisenberg-Weyl algebra~\cite{fein} the NC space structure abides the commutation relation
\begin{equation}\label{b1}\nonumber
[x_{i},x_{j}]=i\theta_{ij}, \,\,\, [x_{i},p_{j}]= i\hbar \delta_{ij}, \,\, [p_{i},p_{j}]=0, 
\end{equation}
where $\theta_{ij}=\epsilon_{ij}\theta$ is the non-commutative parameter and $\delta_{ij}$ is the Kronecker delta which results to one when $i=j$ and zero otherwise. Here $\epsilon_{ij}$ is an antisymmetric matrix and so the non-commutative parameter $\theta_{ij}$ is a real and anti-symmetric matrix. The non-commutative parameter in the space-space like case, i.e., when the space coordinate and the time commutes with each other, the dimension of $\theta_{ij}$ is $(length)^2$ and in the case of space-time, the dimension of this parameter is $length\, . \,time$. In our space structure models, the non-commutative parameters and their associated fundamental lengths are of the order of the Planck length. In this work, we have studied the models and their effects on thermodynamic cycles in natural units where $\hbar = 1$ and $c=1$. One can derive this relation by using the star product definition 
\begin{equation}\label{b2} 
 f(x) \star h(x)=\exp\left\{{i\over 2}\theta_{ij} \partial_{x^{i}}\partial_{y^{j}}\right\} f(x)h(y){\Big{|}}_{x=y},
\end{equation}
where $f$ and $h$ are two arbitrary functions of two variables $x$ and $y$. This defines the generalized quantum mechanics which boils down to the standard one when $\theta=0$. We can develop the Hamiltonian for this space structure by using the definition~\eqref{b2} in Eq.~\eqref{a4}, for NC space structure. The Hamiltonian takes the form 
\begin{eqnarray}\label{b3} \nonumber
H_{\alpha_1}^{NC} & = & {1\over 2m}\left(p^{2}_{1} + p^{2}_{2}
\right) + {c_1\over 2} \left(x_{1}-{\theta \over 2\hbar}p_2 \right)^{2} \\ \nonumber
& + & {c_2\over 2} \left(x_{2}+{\theta \over 2\hbar}p_1 \right)^{2} \\
& + & {c_3\over 2} \left(x_{1}-{\theta \over 2\hbar}p_2\right )
\left(x_{2}+{\theta \over 2\hbar}p_1\right ). 
\end{eqnarray}

By transforming the Hamiltonian~\eqref{b3}, we can develop a compact form for the Hamiltonian. It takes the form 
\begin{eqnarray}\label{b4} \nonumber
  H_1^{NC} & = & {1\over 2M}\left({\Xi}_{1}^2 +
{\Xi}_{2}^2\right) + {K\over 2} \left( \Theta_1^2 + \Theta_2^2 \right) \\
& + & {K\theta \over 2\hbar} \Big(\Theta_2 \Xi_1 - \Theta_1 \Xi_2\Big),
\end{eqnarray} 
where $M$ depicts the effective mass of the system. It is described as $M = {m\over 1 +\left({m\omega_\vartheta \theta \over 2\hbar}\right)^2}$. The effective mass, $M$, boils down to the defined mass $m$ when $\theta = 0$. To establish the compact form of the Hamiltonian we have rescaled the variables to new co-ordinates $\Xi_i$ and $\Theta_i$. If we compare Eq.~\eqref{b4} with Eq.~\eqref{a7}, we will encounter an extra term in the Hamiltonian of NC space which is a function of $\theta$. The new co-ordinates defined in Eq.~\eqref{b4} ($\Theta$, $\Xi$) which represents the position and the momentum variables respectively are expressed in terms of creation and annihilation operators as 
\begin{equation} \label{b5}
\Theta_i =\sqrt{\hbar\Omega\over 2K} \left(b_i+b_i^{\dag} \right),\quad \Xi_i = i
\sqrt{M\hbar\Omega\over 2} \left(b_i^{\dag} -b_i\right),
\end{equation}
which satisfy the relation
\begin{equation}\label{b6} \nonumber
[b_i, b_j^{\dag}] = \delta_{ij},
\end{equation}
where $b_i, b_j^{\dag}$ are the annihilation and creation operator respectively.
The effective frequency $\Omega$ is a function of $\theta$ and is described as $\Omega = \sqrt{K\over M}$. We can re-define the Hamiltonian in Eq.~\eqref{b4} by transforming it with another set of operators. The new Hamiltonian takes the form 
\begin{equation}\label{b7}
H_{final}^{NC} = \hbar \Omega_1 B_1^{\dag}B_1 +
\hbar \Omega_2 B_2^{\dag}B_2 + \hbar \Omega,
\end{equation}
where $\Omega_1 = \Omega + {K \theta \over 2\hbar}, \Omega_2 = \Omega - {K\theta \over 2\hbar}$. The new operators are expressed as
\begin{eqnarray}\label{b8}\nonumber
B_1  =  {1\over \sqrt{2}} (b_1+ib_2), \quad B_2 = {1\over \sqrt{2}} (-b_1+ib_2),\\ \nonumber
B_1 = {1\over \sqrt{2}} (b_1^{\dag}-ib_2^{\dag}), \quad B_2 = {1\over \sqrt{2}} (-b_1^{\dag}-ib_2^{\dag}),
\end{eqnarray}
where $b_1, b_2$ are the annihilation operator and $b_1^{\dag}, b_2^{\dag}$ are the creation operator.
The most compact form of the Hamiltonian for the considered NC space structure is defined in Eq.~\eqref{b7}. Solving this Hamiltonian the energy eigenvalues results as 
\begin{equation}\label{b9} 
E_{n_1,n_2}^{NC} = \hbar \Omega_1 n_1 +\hbar \Omega_2 n_2+ \hbar \Omega.
\end{equation}
The energy spectrum for the coupled HO for the considered NC space structure is depicted in the Eq.~\eqref{b9}.

%%%%%%%%%%%%%%%%%%%%%%%%%%%%%%%%%%%%%%%%%%%%%%%%%%%%%%%%%%%%%%%%%%%%

\subsection{Coupled HO for generalized NC space}\label{sec2b}
We will consider a generalized NC space for our analysis. We call this generalized NC space because the deformation is considered for both the co-ordinate and momentum space, i.e., the commutation relation for both these space structure results to non-zero. The position and the momentum of this  space structure satisfies the following commutation relation 
\begin{equation}\label{c1} \nonumber
    [\hat{x}_{i},\hat{x}_{j}]= i\gamma_{ij}~,
    \quad[\hat{p}_{i},\hat{p}_{j}]=i \xi_{ij}~,
    \quad[\hat{x}_{i},\hat{p}_{j}]= i \hbar \delta_{ij}~,
\end{equation}
where $\gamma_{ij}= \epsilon_{ij} \gamma$, $\xi_{ij} = \epsilon_{ij} \xi$ is  the non-commutative parameter and $\delta_{ij}$ represents the Kronecker delta. Here $\epsilon_{ij}$ represents an antisymmetric matrix. We can define the Hamiltonian for this space structure (as in \cite{linb}) by separating  out the Hamiltonian~\eqref{a7} into two parts which are described as 
\begin{eqnarray}\label{c2} \nonumber
\mathcal{H}_1^{GNC} & = & \left(\frac{e^{\zeta}\sqrt{K}\sin
a}{\sqrt{2}}~y_\alpha+\frac{\cos a}{\sqrt{2m}}~q_ \varpi\right)^2 \\ \nonumber
& + & \left(\frac{e^{-\zeta}\sqrt{K}\sin b}{\sqrt{2}}~y_ \varpi+\frac{\cos
b}{\sqrt{2m}}~q_\alpha\right)^2,\\ \nonumber
\\ \nonumber
\mathcal{H}_2^{GNC} & = & \left(\frac{e^{\zeta}\sqrt{K}\cos
a}{\sqrt{2}}~y_\alpha-\frac{\sin a}{\sqrt{2m}}~q_ \varpi\right)^2 \\
& + & \left(\frac{e^{-\zeta}\sqrt{K}\cos b}{\sqrt{2}}~y_ \varpi-\frac{\sin
b}{\sqrt{2m}}~q_\alpha\right)^2.\,\,\,\,\,\,\,
\end{eqnarray}

Here $a$ and $b$ take values such that 
\begin{equation}\label{c3} \nonumber
\begin{split}
    \sin(a-b)&=\frac{\hbar\sqrt{Km}(e^{\zeta}+e^{-\zeta})}{\lambda_1},
    \cos(a-b)=\frac{Km\gamma-\xi}{\lambda_1},\\
    \sin(a+b)&=\frac{\hbar\sqrt{Km}(e^{\zeta}-e^{-\zeta})}{\lambda_2},
    \cos(a+b)=-\frac{Km\gamma+\xi}{\lambda_2},
\end{split}
\end{equation}
where $ \lambda_1=(e^{\zeta}+e^{-\zeta})\hbar\sqrt{Km}\sqrt{1+\Delta_1}~$, and  $\lambda_2=(e^{\zeta}-e^{-\zeta})\hbar\sqrt{Km}\sqrt{1+\Delta_2}~$. Here $\Delta_1$ and $\Delta_2$ denote the non-commutative effect of the phase space. When $\gamma= \xi=0$, we have $\Delta_1=\Delta_2=0$, and it returns to the ordinary commutative phase space. $\Delta_1,~\Delta_2$ is evaluated as $ \Delta_1=\frac{(Km\gamma-\xi)^2}{(e^{\zeta}+e^{-\zeta})^{2}\hbar^2Km}~,$ and $\Delta_2=\frac{(Km\gamma+\xi)^2}{(e^{\zeta}-e^{-\zeta})^{2}\hbar^2Km}$.

By further simplification the compact form for $a$ and $b$ is evaluated as 
\begin{eqnarray}\label{c4} \nonumber
    a & = & \frac{1}{2}\Big(\arctan \Lambda + \arctan \varkappa \Big),\\ \nonumber
    b & = & \frac{1}{2}\Big(\arctan \Lambda - \arctan \varkappa \Big),
\end{eqnarray}

where $\Lambda = \frac{\hbar\sqrt{Km}(e^{\zeta}-e^{-\zeta})}{-(Km\gamma+\xi)} $, and $\varkappa = \frac{\hbar\sqrt{Km}(e^{\zeta}+e^{-\zeta})}{Km\gamma-\xi}$. To obtain the eigenvalues we have to solve the Hamiltonian~\eqref{c2}. So, the energy eigenvalues of the Hamiltonian for the considered NC space results to 
\begin{eqnarray}\label{c5}
E_{n_1,n_2}&=& E^{^{(1)}}_{n_1}+E^{^{(2)}}_{n_2}\nonumber\\
    &=&\frac{1}{2m}\Big((n_1+n_2+1)\lambda_1+(n_1-n_2)\lambda_2\Big)\nonumber\\
    &=&\frac{\hbar\omega_\vartheta}{2}\Big((n_1+n_2+1)(e^{\zeta}+e^{-\zeta})\sqrt{1+\Delta_1}  \nonumber\\
  & + & (n_1-n_2)(e^{\zeta}-e^{-\zeta})\sqrt{1+\Delta_2}~\Big)\nonumber\\ 
    &=& \hbar \Big(n_1 \omega_a + n_2 \omega_b + {\omega \over 2} (e^{\zeta}+e^{-\zeta})\sqrt{1+\Delta_1} \Big),\nonumber\\
\end{eqnarray}

where the frequencies are defined as 
\begin{eqnarray}\label{c6} \nonumber
\omega_a & = & \frac{\omega_\vartheta}{2} \Big\{ (e^{\zeta} + e^{-\zeta}) \sqrt{1+\Delta_1} + (e^{\zeta} - e^{-\zeta}) \sqrt{1+\Delta_1} \Big\}\\\nonumber
\omega_b & = & \frac{\omega_\vartheta}{2} \Big\{ (e^{\zeta} + e^{-\zeta}) \sqrt{1+\Delta_1} - (e^{\zeta} - e^{-\zeta}) \sqrt{1+\Delta_1} \Big\}.\\
\end{eqnarray}
 So, the energy spectrum of coupled harmonic oscillator for the generalized NC space is depicted in Eq.~\eqref{c5}.

%%%%%%%%%%%%%%%%%%%%%%%%%%%%%%%%%%%%%%%%%%%%%%%%%%%%%%%%%%%%%%%%%%%%%%%%%%%%%%%%%%%%

\section{Otto cycle with coupled harmonic oscillator}\label{sec3}
We will briefly describe the quantum Otto engine proposed by Kieu~\cite{tdk}. The Otto cycle in the classical realm is composed of two adiabatic processes and two isochoric processes, which are being used in automobile piston engines. In the classical engine model  the adiabatic processes are described by the adiabatic expansion and contraction of the piston. In the quantum regime, the engine cycles are modeled based on different quantum systems as its working medium instead of considering an ideal gas. For our analysis, we have adopted a quantum system, namely, coupled harmonic oscillator as the working substance for the engine cycle. The adiabatic increase and decrease in the quantum realm are controlled by the change in the energy levels which occurs due to the change in the frequency of the oscillator. The isochoric process in the quantum version is represented by the thermalization processes, during which it exchanges heat with the bath. Work is done during the adiabatic process of the cycle. The change in the mean energies guides us to calculate the work and heat for the cycle, where mean energy for the system is represented in terms of the state $\rho$ and the Hamiltonian $H$. It is defined as $Tr[\rho H]$.

\begin{figure}[h]
\center
  \includegraphics[width=1.0\columnwidth]{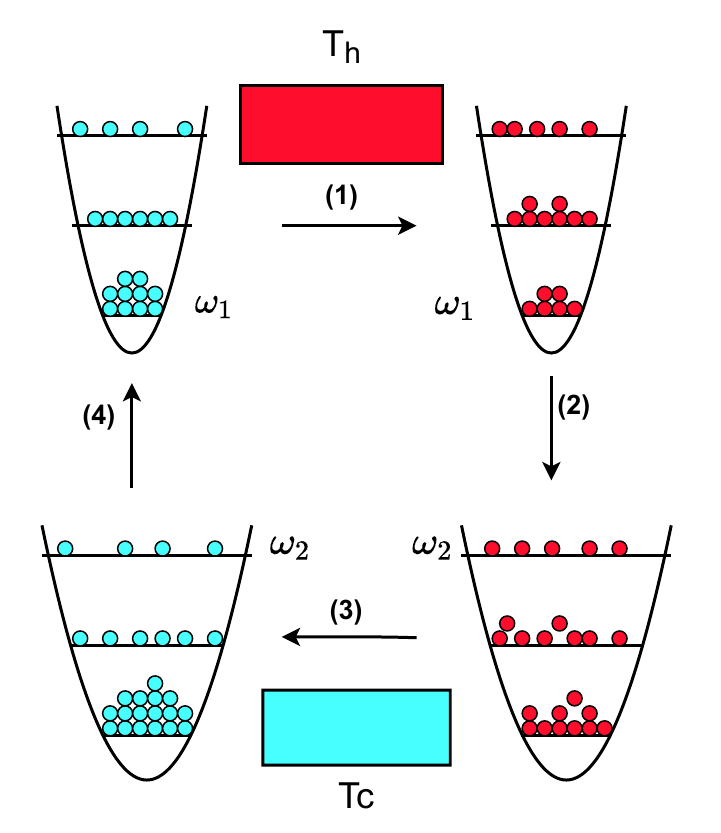}
  \caption{(Color online) The schematic representation of a quantum Otto cycle is shown.  The working substance of the cycle is a coupled harmonic oscillator. The first stage and the third stage of the cycle are the thermalization processes, and the second phase and the fourth phase corresponds to the adiabatic processes.}
  \label{figa}
  \end{figure}

The four-staged of a quantum Otto cycle~\cite{thom} with harmonic oscillator as the working medium is schematically described in Fig.~(\ref{figa}). The four stages of the quantum Otto cycle are

%\begin{figure}[h]
%\center
%  \includegraphics[width=1.0\columnwidth]{Otto_cycle.pdf}
%  \caption{Schematic representation of the Otto cycle (T-S diagram) in classical regime, where $AB$ and $CD$ describes the isochoric processes and $BC$ and $DA$ the adiabatic process of the cycle.}
%  \label{fig123}
%  \end{figure}

(1)  The first stage is the isochoric process. During this stage of the cycle, the working medium is coupled with the bath at temperature $T_h$. In other words, the Hamiltonian $H^{(1)}$ is attached to the hot bath. The system is represented by the density matrix $\rho_c^{(2)}$. The Hamiltonian is fixed throughout this process. The system approaches equilibrium with the bath at the end of this process. So, the final state of the system after this  stage is given as $\rho_h^{(1)} = \frac{exp(-\beta_h H)}{Tr[exp(-\beta_h H)]}$, where $\beta_h=1/k_BT_h$, with $k_B$ as the Boltzmann constant. The amount of heat absorbed from the bath at temperature $T_h$ is $Q_{hot} = Tr[H(\rho_h^{(1)} - \rho_c^{(2)})]$. 

(2) The second stage of the cycle is the adiabatic process. The working medium in this phase is thermally isolated so that the quantum adiabatic theorem is valid throughout the process. The Hamiltonian of the system changes from $H^{(1)}$ to $H^{(2)}$. During this process, we do not encounter any heat exchange between the system and the bath. So, the change in energy is equivalent to the work done. The work done is described as 
\begin{equation}\label{d1} \nonumber
W_1 = Tr [(\rho_h^{(1)} H^{(1)} - \rho_h^{(2)} H^{(2)})],
\end{equation}
where $\rho_h^{(2)} = U_1 \rho_h^{(1)} U_1^\dag$, with $U_1$ as the unitary operator which is associated with the adiabatic process. It is defined as 
\begin{equation}\label{d2} \nonumber
U_1 = \mathcal{T} e^{[-(\frac{i}{\hbar}) \int_{0}^{\mathcal{T}} H(t) dt]},
\end{equation}
where $\mathcal{T}$ represents the total time of evolution for the quasi-static process. Here $H(0) = H^{(1)}$ and $H(\mathcal{T})= H^{(2)}$.

\textcolor{black}{We know that adiabaticity can be understood from the driving frequency of our system. For our analysis, we took the cycles slow enough to ensure the adiabaticity condition.  For the analysis of our work, we have assumed that the spectrum of the system lies in the  NC space, whereas our bath happens to be a conventional bath that is generally used. So, in this case, we can see adiabaticity has nothing to do with NC space structure.} \textcolor{black}{For our analysis, we have not considered any type of correlation. Now we know that in non-commutative space, one can expect an increase in the entanglement as it provides an extra degree of freedom than the usual space~\cite{deyyy}, and it is well known that correlation helps in better work extraction. So, we can expect that the correlation will boost the performance of the cycle. }

(3) The third stage is represented by the isochoric process. At this phase of the cycle the system is coupled with a cold bath at a temperature $T_c$. Similar to the stage one, the system attains equilibrium with the cold bath at the end of this stage. The state of the system at this phase is described as $\rho_c^{(1)} =  \frac{e^{(- \beta_c H^{(2)})}}{Tr[e^{(- \beta_c H^{(2)})}]}$. So, heat is rejected to the bath and is evaluated as $Q_{cold}=Tr[H^{(2)} (\rho_c^{(1)} - \rho_h^{(2)})]$.

(4) The last stage of the cycle is the adiabatic process. In this last phase of the cycle, the working substance is thermally isolated from the reservoir and at the end of this process, the system gets coupled with the hot bath. During this process, the Hamiltonian changes from $H^{(2)}$ to $H^{(1)}$. So, the work done in this process is 
\begin{equation}\label{d3} \nonumber
W_2 = Tr[\rho_c^{(1)} H^{(2)} - \rho_c^{(2)} H^{(1)}],
\end{equation}
which is equivalent to variation in the mean energy. Here $\rho_c^{(2)}$ is the density state of the system at the end of this process. It is defined as $\rho_c^{(2)} = U_2 \rho_c^{(1)} U_2^{\dagger}$, where $U_2$ is evaluated as 
\begin{equation}\label{d4} \nonumber
U_2 = \mathcal{T} e^{[-(\frac{i}{\hbar}) \int_{0}^{\mathcal{T}} H(t) dt]}.
\end{equation}
Here $H(0) = H^{(2)}$ and $H(\mathcal{T})= H^{(1)}$.

During the execution of the process of the Otto cycle, the Hamiltonian of the system evolves from $H(0)= H^{(1)}$ to $H(\mathcal{T})= H^{(2)}$. One can visualize that the NC space follows the standard master equation for the analysis of the evolution of the system as suggested by the previous analysis~\cite{dias1,abcd1234}.

%%%%%%%%%%%%%%%%%%%%%%%%%%%%%%%%%%%%%%%%%%%%%%%%%%%%%%%%%%%%%%%%%%%%%%%%

\subsection{In commutative phase space}\label{sec3a}
Here, we will consider two coupled oscillators as our working substance. The Hamiltonian for this coupled system in commutative phase space is described in Eq.~\eqref{b7}. Now we will consider the Otto cycle described above with the coupled system as the working medium. 

During the first adiabatic process, the Hamiltonian of the working substance of the Otto cycle changes its initial value form $H^{(1)}$ to $H^{(2)}$. The change in the Hamiltonian is due to the change in the eigen-frequency of the oscillators from $\omega_1$ to $\omega_2$. It reverts to its respective initial values after the execution of the second adiabatic process of the cycle.  The total work done by the system is the sum of the contribution of the two oscillators. So, the work done is a function of the frequency of the oscillators and the coupling strength of the two oscillators. The frequency of the two oscillators is considered to be the same for our analysis. During the execution of the adiabatic process, it is assumed that there is no cross over of the energy level of the Hamiltonian of the coupled oscillator. It is also taken care of that the system abides the quantum adiabatic theorem. So, the process occurs slowly enough such that the population of the eigenstate of the Hamiltonian remains constant throughout the process. The total amount of heat absorbed by the working medium from the hot bath is given by
\begin{eqnarray}\nonumber \label{e1}
Q & = & Tr[H(\rho_h^{(1)} - \rho_c^{(2)})] \\ \nonumber
& = & \frac{\hbar \omega_1 e^{\zeta}}{2} \Big( \coth \Big[\frac{\beta_h \hbar \omega_1 e^{\zeta}}{2} \Big] - \coth \Big[\frac{\beta_c \hbar \omega_2 e^{\zeta}}{2} \Big] \Big) \\ \nonumber
& + & \frac{\hbar \omega_1 e^{-\zeta}}{2} \Big( \coth \Big[\frac{\beta_h \hbar \omega_1 e^{-\zeta}}{2} \Big] - \coth \Big[\frac{\beta_c \hbar \omega_2 e^{-\zeta}}{2} \Big] \Big). \\
\end{eqnarray}

The total work done by the Otto cycle is define $W = W_1 + W_2$. So, the work done by the Otto cycle with coupled HO as the working medium is expressed as 
\begin{eqnarray} \nonumber \label{e2}
W & = & \frac{\hbar (\omega_1 - \omega_2) e^{\zeta}}{2} \Big( \coth \Big[\frac{\beta_h \hbar \omega_1 e^{\zeta}}{2} \Big] - \coth \Big[\frac{\beta_c \hbar \omega_2 e^{\zeta}}{2} \Big] \Big)\\ \nonumber
& + & \frac{\hbar (\omega_1 - \omega_2) e^{-\zeta}}{2} \Big( \coth \Big[\frac{\beta_h \hbar \omega_1 e^{-\zeta}}{2} \Big] \\ \nonumber
& - & \coth \Big[\frac{\beta_c \hbar \omega_2 e^{-\zeta}}{2} \Big] \Big).\\
\end{eqnarray}

The efficiency of the coupled system which is considered as the working medium is defined as the ratio of total work done over the total heat absorbed by this system during the execution of the process. It is given as 
\begin{equation}\nonumber \label{e3}
\eta_{otto_C} = \frac{W}{Q} = f(\zeta).
\end{equation}

%\begin{figure}[h]
%\center
%  \includegraphics[width=1.0\columnwidth]{Coupled_p.pdf}
%  \caption{(Color online) Efficiency of the Otto cycle as a function of the coupling parameter in commutative space with coupled HO as the working substance.}
%  \label{fig1}
%  \end{figure}
We have considered the hot reservoir temperature $T_h = 4K$ and that of the cold reservoir temperature $T_c = 1K$. The frequency of the coupled oscillator is considered as $\omega_1= 4$ \textcolor{black}{MHz} and $\omega_2= 3$ \textcolor{black}{MHz} for the evaluation of the efficiency of the Otto cycle with respect to the variation of the coupling strength. For the analysis of the efficiency of the engine we have numerically simulated the efficiency factor with respect to the coupling parameter. The efficiency of the Otto cycle with coupled harmonic oscillator as the working substance remains constant with the change in the coupling parameter. The efficiency for the engine model with the coupled harmonic oscillator considered in our analysis produces an equivalent result to the previously studied models~\cite{thom}. One can infer from all these different analysis that the coupling of harmonic oscillators does not provide any good advantage to the efficiency for this specific engine model. So, we can conclude that the coupling strength of the coupled oscillators results in a constant efficiency even for the two different approaches.

%%%%%%%%%%%%%%%%%%%%%%%%%%%%%%%%%%%%%%%%%%%%%%%%%%%%%%%%%%%%%%%%%%%%%%%%%%%

\subsection{In non-commutative phase space}\label{sec3b}
In the case of commutative space, the coupled oscillators as the working substances result in a constant efficiency with respect to the coupling strength of the system. Now here, we will analyze how the change in the phase space affects the thermodynamic process. The Hamiltonian of the two harmonic oscillators coupled with each other in the NC phase space is described in Eq.~\eqref{b7}. Following the same methodology as used in the case of commutative space, we will analyze the quantum Otto cycle in the NC phase space. We have considered even in NC space that the working medium will evolve to a Gibbs state when coupled to a heat bath similar to the previous analysis in this direction~\cite{girotti}. We have followed this throughout our analysis.

Similar to the case in commutative space, the Hamiltonian changes its initial value from $H^{(1)}$ to $H^{(2)}$ during the first adiabatic process. The change in the Hamiltonian is due to the change in the eigen frequency of the oscillators from $\omega_1$ to $\omega_2$. It goes back to its respective initial values after the second adiabatic process of the cycle. So, the work done is a function of the frequency of the oscillators, the coupling strength of the two oscillators and the NC parameter of the phase space. For our analysis, the frequency of the two oscillators is considered to be the same throughout. We take care of the fact that there is no cross over of the energy levels of the Hamiltonian which ultimately satisfies the quantum adiabatic theorem in the NC phase space. The net amount of heat absorbed is given as
\begin{eqnarray}\nonumber \label{f1}
Q & = & Tr[H(\rho_h^{(1)} - \rho_c^{(2)})] \\ \nonumber
& = & \frac{\hbar \Big(\omega_1  + \frac{K \theta}{2 \hbar} \Big)}{2} \Bigg( \coth \Bigg(\frac{ \beta_h \hbar \Big(\omega_1  + \frac{K \theta}{2 \hbar} \Big)}{2} \Bigg) \\ \nonumber
& - & \coth \Bigg(\frac{ \beta_c \hbar \Big(\omega_2  + \frac{K \theta}{2 \hbar} \Big)}{2} \Bigg) \Bigg) \\ \nonumber
 & + &  \frac{\hbar \Big(\omega_1  - \frac{K \theta}{2 \hbar} \Big)}{2}
   \Bigg( \coth \Bigg(\frac{ \beta_h \hbar \Big(\omega_1  - \frac{K \theta}{2 \hbar} \Big)}{2} \Bigg)  \\ 
& - & \coth \Bigg(\frac{ \beta_c \hbar \Big(\omega_2  - \frac{K \theta}{2 \hbar} \Big)}{2} \Bigg) \Bigg). \quad \quad
\end{eqnarray}

The total work done by the Otto cycle is define as $W = W_1 + W_2$. So, the work done by the Otto cycle with coupled HO as the working medium in NC space is expressed as
\begin{eqnarray}\nonumber \label{f2}
W & = & \frac{\hbar \Big(\omega_1 - \omega_2 \Big)}{2} \Bigg( \coth \Bigg(\frac{ \beta_h \hbar \Big(\omega_1  + \frac{K \theta}{2 \hbar} \Big)}{2} \Bigg) \\ \nonumber
& - & \coth \Bigg(\frac{ \beta_c \hbar \Big(\omega_2  + \frac{K \theta}{2 \hbar} \Big)}{2} \Bigg) \Bigg)\\ \nonumber
 & + &  \frac{\hbar \Big(\omega_1 - \omega_2 \Big)}{2} 
\Bigg( \coth \Bigg(\frac{ \beta_h \hbar \Big(\omega_1  - \frac{K \theta}{2 \hbar} \Big)}{2} \Bigg) \\
& - & \coth \Bigg(\frac{ \beta_c \hbar \Big(\omega_2  - \frac{K \theta}{2 \hbar} \Big)}{2} \Bigg) \Bigg). \quad \quad  
\end{eqnarray}

The efficiency of the coupled system is defined as the ratio of total work done over the total heat absorbed by the system during the execution of the process. It is described as 
\begin{equation}\nonumber \label{f3}
\eta_{otto_{NC}} = \frac{W}{Q} = f(\theta).
\end{equation}

\begin{figure}[h]
\center
  \includegraphics[width=1.0\columnwidth]{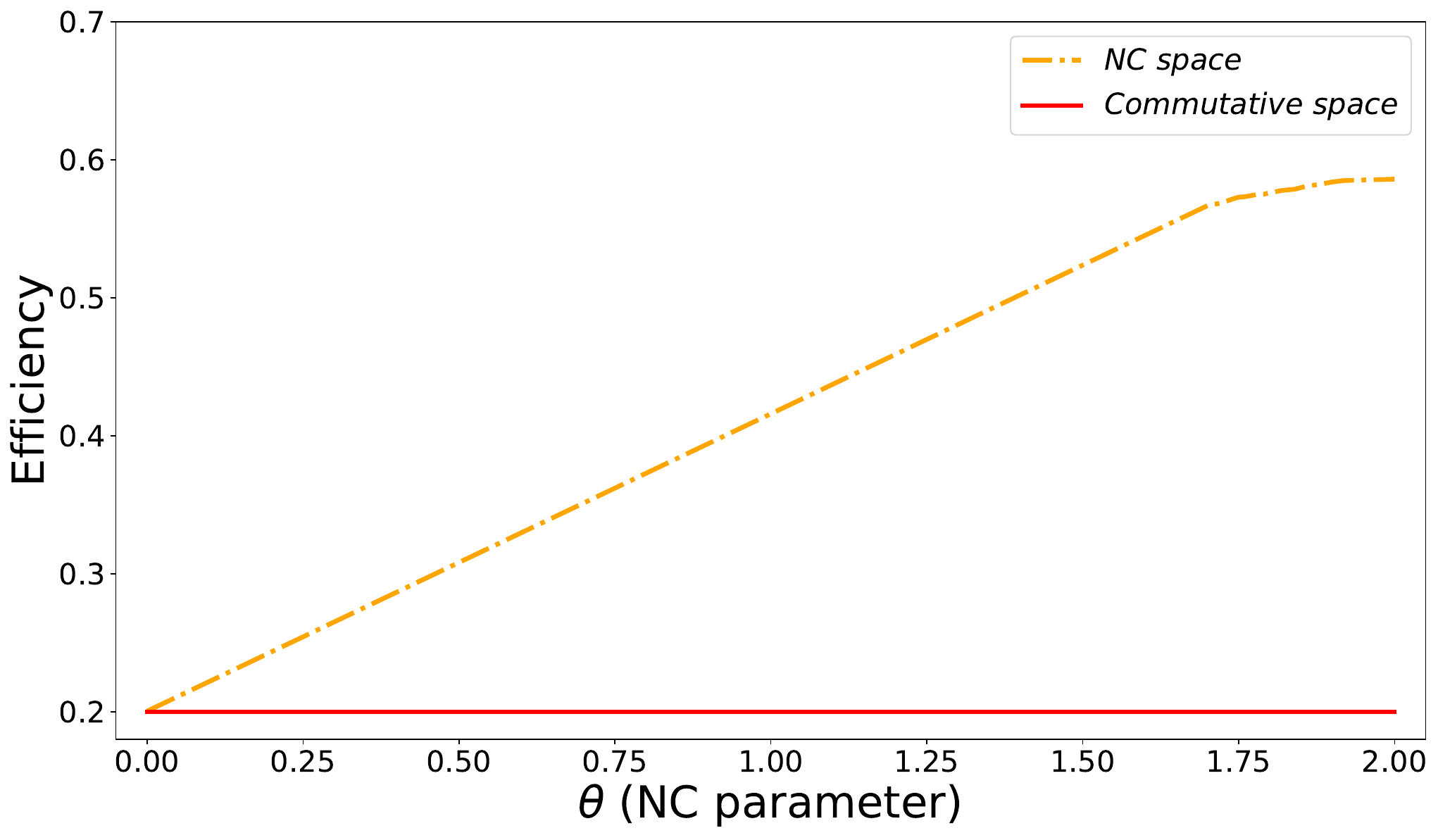}
  \caption{(Color online) Efficiency of the Otto cycle as a function of the NC parameter with coupled HO as the working substance is shown. The orange dash dotted curve depicts the variation of the efficiency with respect to the NC parameter with a constant coupling factor. The red solid line depicts the efficiency of the commutative space with coupled HO as the working substance where the coupling constant $\zeta=2$.} 
  \label{fig2}
  \end{figure}

Similar to the case of commutative space, we have considered the hot reservoir temperature $T_h = 4K$ and that of the cold reservoir temperature $T_c = 1K$. The frequency for the coupled oscillator is considered as $\omega_1= 4$ \textcolor{black}{MHz} and $\omega_2= 3$ \textcolor{black}{MHz} for the evaluation of the efficiency of the Otto cycle with respect to the variation of the NC space parameter. For our analysis, we have fixed the coupling strength $\zeta = 2$ and the constant $K= 0.25$ and \textcolor{black}{$\theta \leq 2\times 10^{-40} m^2$ }. Here we have numerically simulated the efficiency of the engine model in NC space structure with respect to the NC parameter. We can observe that the efficiency of the engine model (as shown in Fig.~(\ref{fig2})) boils down to the efficiency of the commutative space when the NC parameter $\theta$ is close to zero. This satisfies the condition that the results produced by the system reduce to the commutative space when $\theta \rightarrow 0$. With the variation of the NC parameter, we observe that the efficiency increases monotonously for a certain range of the NC parameter. Then at a certain stage the efficiency of the engine gets saturated with the variation of NC parameter. So, we can infer that NC space provides a boost to the efficiency  of the engine over the efficiency of the commutative space.

%%%%%%%%%%%%%%%%%%%%%%%%%%%%%%%%%%%%%%%%%%%%%%%%%%%%%%%%%%%%%%%%%%%%%%%%%%%

\subsection{In generalized non-commutative phase space}\label{sec3c}
In the case of non-commutative phase space, the coupled oscillators as the working substances result in a boost to the efficiency with respect to the NC space parameter of the system. The efficiency is high for the lower values of the NC parameter. Now, we will analyze how the generalized NC phase space affects the thermodynamic process. The Hamiltonian of the two coupled harmonic oscillators in the NC phase space is described in Eq.~\eqref{c2}. Following the same methodology, as used in the case of non-commutative space we will analyze the quantum Otto cycle in the NC phase space.

In the case of generalized NC phase space, the Hamiltonian is separated into two part as shown in Eq.~\eqref{c2}. During the adiabatic process, the individual Hamiltonian changes from the initial values from $H^{(1)}$ to $H^{(2)}$. The total Hamiltonian is the sum of the effect of these two Hamiltonians. The changes in the Hamiltonian is due to the change in the eigen frequency of the oscillators, where the eigen frequency for the first oscillator changes from $\omega_1$ to $\omega_1^{'}$ and for the second oscillator the frequency changes from $\omega_2$ to $\omega_2^{'}$. After the second adiabatic process, the eigen frequencies return to the respective initial stage. So, we can consider that the working substance is composed of two independent oscillators. The total work done is a result of the contribution of the two oscillators. Therefore, the work done is a function of the frequency of the system and the NC phase space parameters. Similar to the previous cases, we have to take care of the fact that there is no cross over of the energy levels of the total Hamiltonian during the execution of the adiabatic process. The total amount of heat absorbed by the system is described as  
\begin{eqnarray}\nonumber \label{g1}
Q & = & Tr[H(\rho_h^{(1)} - \rho_c^{(2)})] \\ \nonumber
& = & \frac{\omega_1}{2} \Bigg( \coth \Big[\frac{\beta_h \omega_1}{2} \Big] - \coth \Big[\frac{\beta_c \omega_1^{'}}{2} \Big] \Bigg) \\
&  + & \frac{\omega_2}{2} \Bigg( \coth \Big[\frac{\beta_h \omega_2}{2} \Big] - \coth \Big[\frac{ \beta_c \omega_2^{'}}{2} \Big] \Bigg).\quad \quad
\end{eqnarray}

The total work done by the system is equivalent to the sum of the work done by the individual systems. So, the total work done by the system can be expressed as 
\begin{eqnarray}\nonumber \label{g2}
W & = & \frac{(\omega_1 - \omega_1^{'})}{2} \Bigg( \coth \Big[\frac{\beta_h \omega_1}{2} \Big] - \coth \Big[\beta_c \omega_1^{'} \Big] \Bigg) \\ \nonumber
& + & \frac{(\omega_2 - \omega_2^{'})}{2} \Bigg( \coth \Big[\frac{\beta_h \omega_2}{2} \Big] - \coth \Big[\beta_c \omega_2^{'} \Big] \Bigg).\\
\end{eqnarray}
where the frequencies of the system is defined equivalent to the Eq.~\eqref{c6}.

The efficiency of the coupled harmonic oscillator system  for the generalized NC phase space is defined as the ratio of total work done over the total heat absorbed by this system. It is described as 
\begin{equation}\label{g3}
\eta_{otto_{GNC}} = \frac{W}{Q} = f(\gamma,\xi).
\end{equation}

\begin{figure}[h]
\center
  \includegraphics[width=1.0\columnwidth]{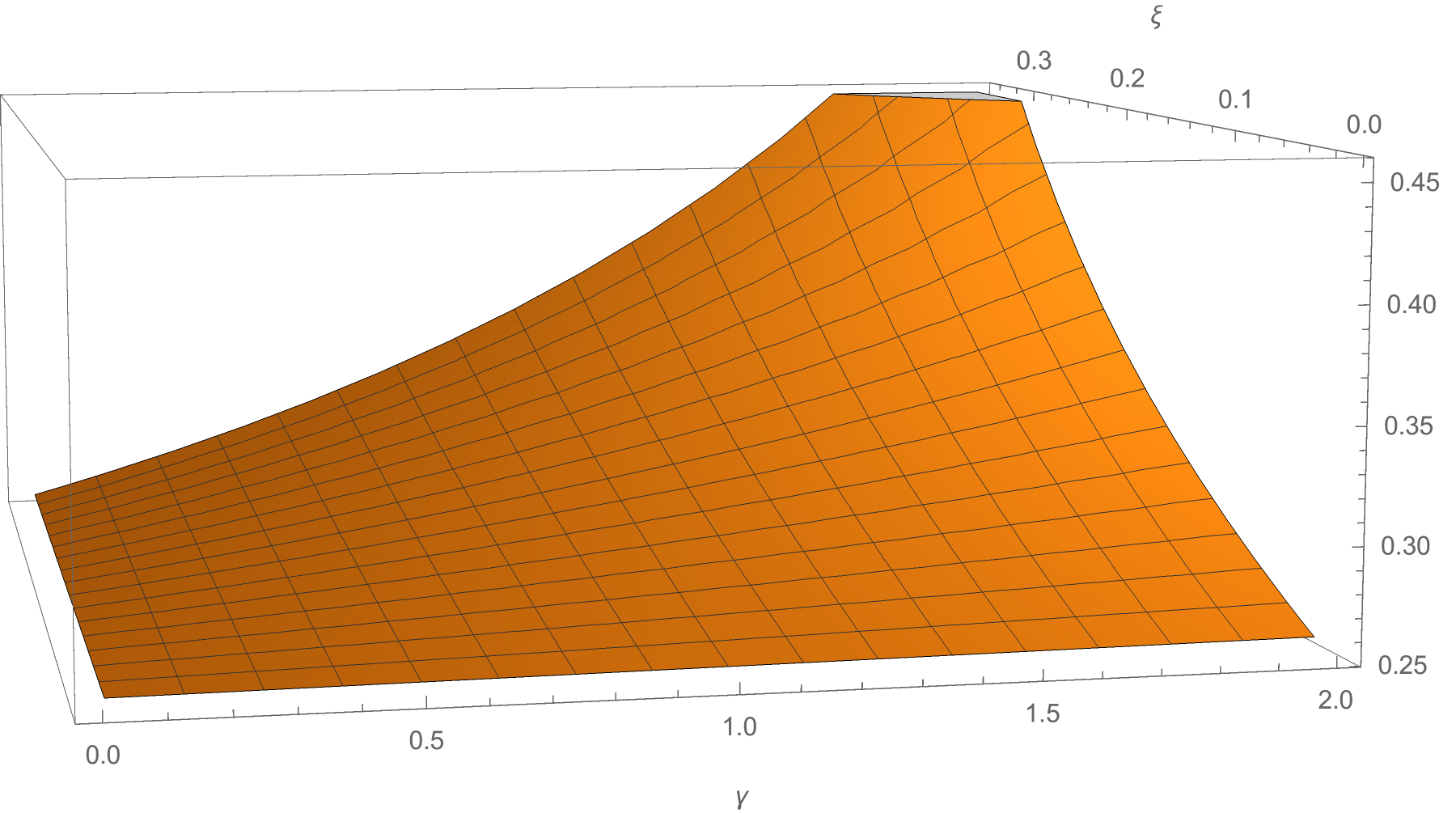}
  \caption{(Color online) Efficiency of the Otto cycle as a function of the NC parameters in generalized NC phase space parameters with coupled HO as the working substance.}
  \label{fig3}
  \end{figure}

Similar to the case of non-commutative space, we have considered the hot reservoir temperature $T_h = 4K$ and that of the cold reservoir temperature $T_c = 1K$ for the analysis of the Otto cycle in generalized NC phase space. Though we define this as generalized NC space, actually we incorporate the deformation in momentum space as well as the coordinate space. So, in this space structure, the NC effect increases than the above one. The frequency for the coupled oscillator is taken as $\omega_1= 4$ \textcolor{black}{MHz} and $\omega_2= 3$ \textcolor{black}{MHz} for the evaluation of the efficiency of the Otto cycle with respect to the variation of the generalized NC space parameters. The coupling strength is taken as $\zeta = 2$ and the constant $K= 0.25$ throughout the process. The NC parameter \textcolor{black}{$\gamma \leq 2\times 10^{-40}m^2$ } and \textcolor{black}{$\xi \leq 2 \times 10^{-61}m^2/s^2$} are bounded within this values. The three-dimensional plot (in Fig.~(\ref{fig3})) shows the variation of the efficiency of the Otto cycle with coupled harmonic oscillator as the working substance in the generalized NC phase space. The variation of the NC parameter of the coordinate and the momentum space, i.e., $\xi$ and $\gamma$ in the graph shows that it as a direct impact on the efficiency of the engine. So, the so-called generalized NC space results in better efficiency of the engine compared to the NC space considered above for the analysis.

%%%%%%%%%%%%%%%%%%%%%%%%%%%%%%%%%%%%%%%%%%%%%%%%%%%%%%%%%%%%%%%%%%%%%%%%%%

\section{Stirling cycle with coupled harmonic oscillator}\label{sec4}

\begin{figure}[h]
\center
  \includegraphics[width=1.0\columnwidth]{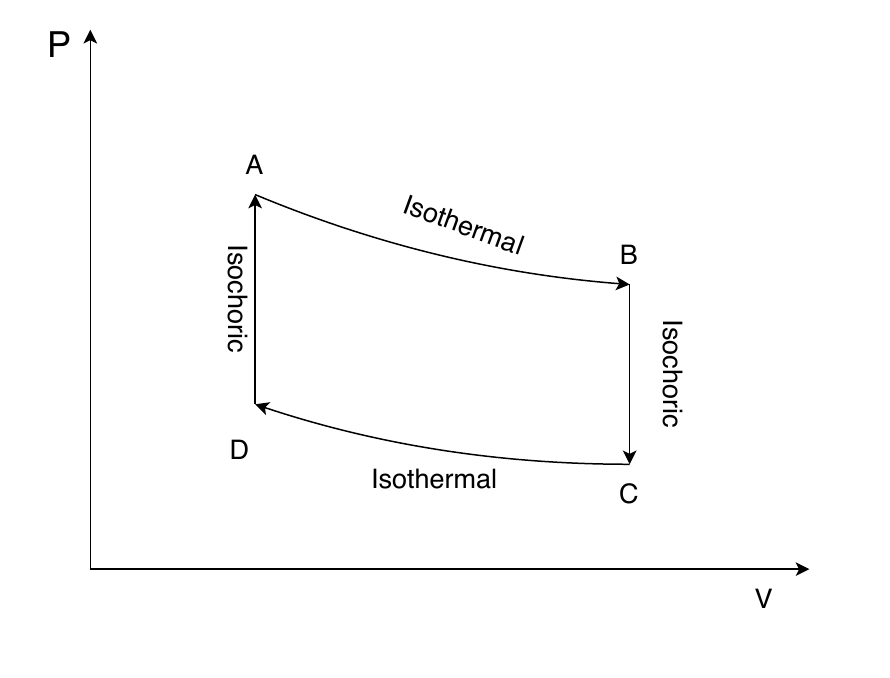}
  \caption{Schematic representation of the Stirling cycle in classical regime, where $AB$ and $CD$ describes the isothermal processes and $BC$ and $DA$ the isochoric process of the cycle. During the $DA$ and $AB$ processes the system remain connected to a hot bath and for the other two processes the system remains coupled with the cold bath.}
  \label{fig4}
  \end{figure}

A Stirling cycle~\cite{sayg,gsaa,xlh,vbli,chatt,chatt1}, which is an example of a discrete engine is a four-stroke engine that comprises of two isothermal processes and two isochoric processes. It is a reversible thermodynamic cycle. The pressure-volume (P-V) diagram of the Stirling cycle in the classical realm is depicted in Fig.~(\ref{fig4}).

\begin{figure}[h]
\center
  \includegraphics[width=1.0\columnwidth]{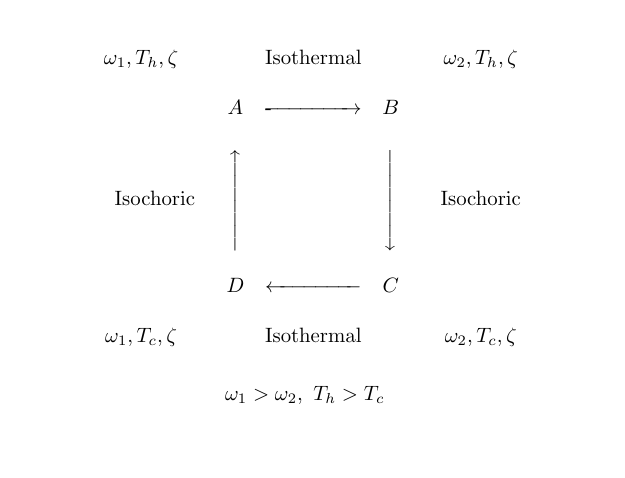}
  \caption{Schematic representation of the Stirling cycle in quantum realm with harmonic oscillator as the working substance. $\zeta$ represents the coupling strength of the coupled harmonic oscillator.}
  \label{fig5}
  \end{figure}

In the quantum realm, the engine is modeled by different working substances like a 1-D potential well, harmonic oscillator, etc. We will analyze the Stirling cycle with coupled HO as the working substance. The isochoric process in the quantum version is represented by the thermalization processes during which it exchanges heat with the bath. The isothermal expansion and compression in the quantum realm are controlled by the change in the energy levels due to the variation in the frequency of the oscillators. The Stirling cycle in a quantum regime with coupled HO as the working substance is depicted in Fig.~(\ref{fig5}). The four stages of the quantum Stirling cycle are

(i)First stage of the Stirling cycle: the \textit{isothermal process} $(A\rightarrow B)$. The working substance in this stage is coupled with the heat bath at temperature $T_h$. All the way round during the execution of this process, the system remains in thermal equilibrium with the hot reservoir. Due to the quasi-static changes in the Hamiltonian of the working medium, we encounter changes in the energy spectrum and the internal energy of the system. During this process, heat is extracted from the bath isothermally.

(ii) Second phase: the \textit{isochoric process} $(B\rightarrow C)$. The system undergoes an isochoric heat exchange while is goes through this phase of the cycle. The system is now decoupled from the hot reservoir and coupled with the cold reservoir at temperature $T_c$. So, heat gets released during this process of the cycle. 

(iii) Third phase: the \textit{isothermal process} $(C\rightarrow D)$. The system remains connected to the cold reservoir at temperature $T_c$ throughout this process. This phase follows the  same condition that is being followed during the execution of the first isothermal process. The system remains at thermal equilibrium with the reservoir. So, heat is rejected to the reservoir in this stage of the  cycle. 

(iv) Fourth stage: the \textit{isochoric process} $(D\rightarrow A)$. The system is decoupled from the cold reservoir and reverted to the hot reservoir at temperature $T_h$. So, heat is extracted from the bath in this process of the cycle.

The efficiency of the Stirling Cycle is defined as the ratio of work output to the heat input. For our analysis, the efficiency  being a function of the coupling strength and the NC parameters in NC phase space.

%%%%%%%%%%%%%%%%%%%%%%%%%%%%%%%%%%%%%%%%%%%%%%%%%%%%%%%%%%%%%%%%%%%%%%%%%%%

\subsection{Commutative phase space}\label{sec4a}

Now, for our analysis, we will consider a coupled HO in commutative phase space as the working substance of the Stirling cycle. The Hamiltonian of the system is described in Eq.~\eqref{a7}. The energy eigenvalues for this Hamiltonian is evaluated as conveyed in Eq.~\eqref{a8}. The partition function for the considered system is described as
 \begin{eqnarray}\label{h1}
 Z & = & \sum_n e^{-\beta E_n}, 
  =  \frac{e^{-\beta \omega \cosh(\zeta)}}{\beta^2 \omega^2}, 
 \end{eqnarray}
where the system has to satisfy the condition $Re \Big[e^{\zeta} \beta \omega \Big]> 0$. 

The heat exchange that takes place during the first stage (i.e, the isothermal process) of the Stirling cycle is
\begin{equation}\label{h2}
Q_{AB}  =  U_A - U_B + \frac{1}{\beta_h} ln \Big( \frac{Z_{\omega_1,\beta_h}}{Z_{\omega_2,\beta_h}} \Big),
\end{equation}
One can evaluate the partition function $Z_A$, $Z_B$ of the system using Eq.~\eqref{h1}. The internal energy $U_A$, $U_B$ is evaluated using the definition $U_i = - \partial ln Z_i/ \partial \beta_h$, where $i= A, \, B$. The internal energy is described as 
 \begin{equation}\nonumber \label{h3}
 U = -\frac{2}{\beta} - \omega \cosh(\zeta).
 \end{equation}

In the second phase of the cycle heat is released from the system. The heat exchange throughout this process can be expressed as
\begin{equation}\label{h4}
Q_{BC} = U_C -U_B.
\end{equation}

The third phase is again a isothermal process. Heat gets rejected from the system in this stage. The heat exchange is represented as
\begin{equation}\label{h5}
Q_{DC}  =  U_D - U_C + \frac{1}{\beta_c} ln \Big( \frac{Z_{\omega_1,\beta_c}}{Z_{\omega_2,\beta_c}} \Big),
\end{equation}

where $U_i = - \partial ln Z_i/ \partial \beta_h$ with $i= C, \, D$.

And in the final stage of the cycle the system undergoes an isochoric heat addition process. So, the heat addition to the system can be expressed as 
\begin{equation}\label{h6}
Q_{DA}  = U_A - U_D.
\end{equation}

The net work done by the cycle is $W_{tot_C} = Q_{AB} + Q_{BC} + Q_{CD} + Q_{DA}$. The efficiency of the Stirling heat cycle from Eq.~\eqref{h2}, ~\eqref{h4}, ~\eqref{h5} and ~\eqref{h6} is defined as 
\begin{eqnarray} \label{h7} \nonumber
\eta_{Stir_C}= 1 + \frac{Q_{BC} + Q_{CD}}{Q_{DA} + Q_{AB}}.
\end{eqnarray}

\begin{figure}[h]
\center
  \includegraphics[width=1.0\columnwidth]{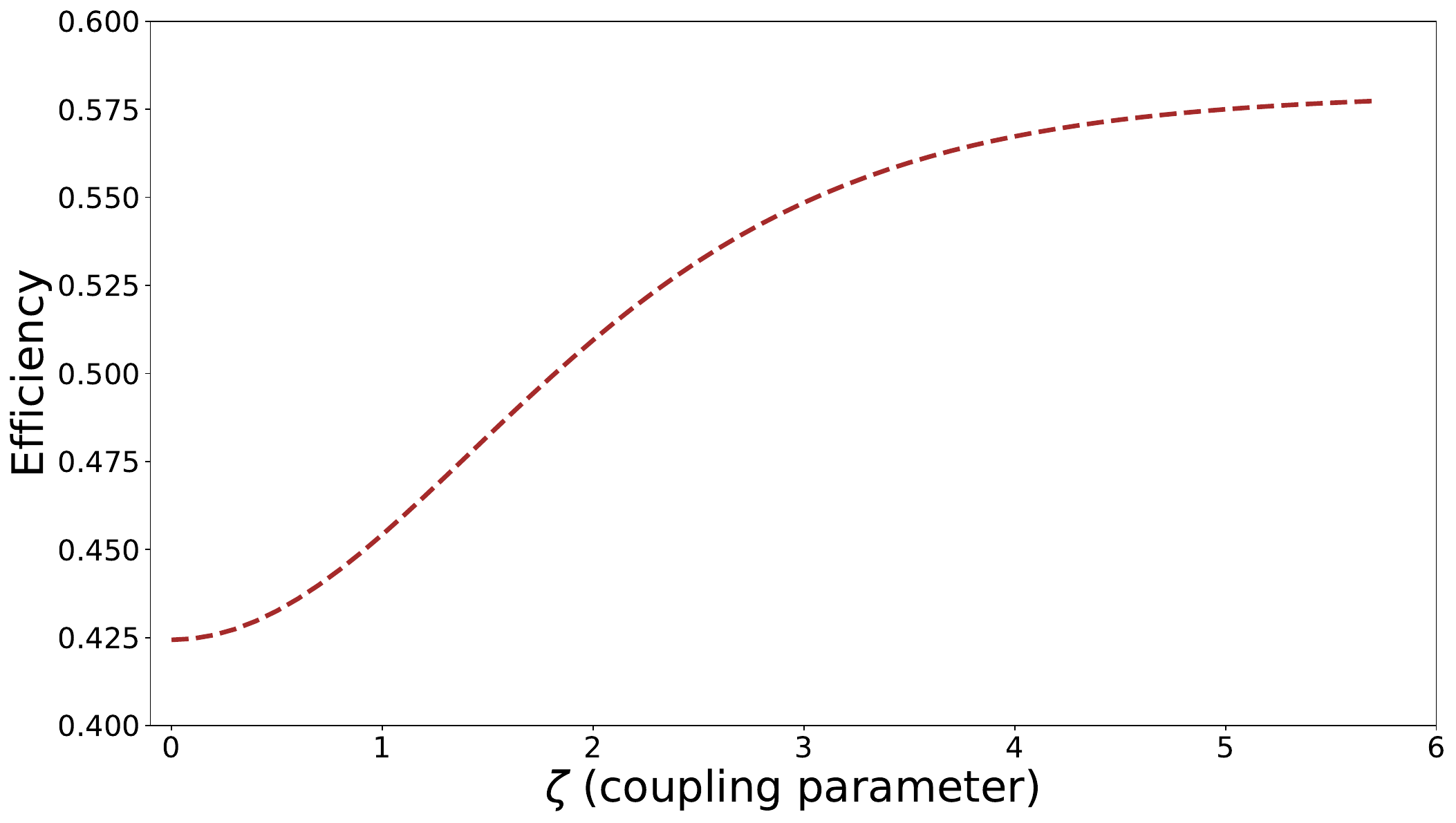}
  \caption{(Color online) Efficiency of the Stirling cycle with coupled HO as the working  substance in commutative phase space.}
  \label{fig6}
  \end{figure}
  
The hot reservoir temperature is considered as $T_h = 4K$, and that of the cold reservoir temperature $T_c = 1K$. For the analysis of the efficiency of the Stirling cycle with respect to the variation of the coupling strength, the frequency of the coupled oscillator is considered as $\omega_1= 4$ \textcolor{black}{MHz} and $\omega_2= 2$ \textcolor{black}{MHz}. We have numerically simulated the efficiency of the Stirling cycle with respect to the coupling constant similar to the Otto cycle studied in the above section. The variation of the efficiency of the engine with respect to the coupling parameter is shown in Fig.~(\ref{fig6}). The efficiency shows very minute variation when the coupling constant is near to zero. The efficiency monotonically increases for a certain range of the coupling constant. Then it gets saturated at a certain value of the coupling strength, and thereafter it remains constant with respect to the parameter. We encounter a high efficiency of the Stirling engine when compared with the Otto cycle with the coupled HO as the working medium.

%%%%%%%%%%%%%%%%%%%%%%%%%%%%%%%%%%%%%%%%%%%%%%%%%%%%%%%%%%%%%%%%%%%%%%%%%%%

\subsection{Non-commutative phase space}\label{sec4b}
In the case of commutative space, the coupled oscillators as the working substances result in a boost to the efficiency with respect to the coupling strength of the system. Now, we want to analyze how the change in the phase space affects the thermodynamic process. The Hamiltonian of the two harmonic oscillators coupled with each other in the NC phase space is described in Eq.~\eqref{b7}. The energy eigenvalue of the Hamiltonian is expressed in the form shown in Eq.~\eqref{b9}. Following the same methodology, as used in the case of commutative space, we will analyze the quantum Stirling cycle in the NC phase space.

The partition function for the system is evaluated and it takes the form 
\begin{eqnarray} \label{i1}
 Z & = & \sum_n e^{-\beta E_n} 
  =  \frac{-4}{\beta^2 \theta^2(K-4\omega^2)},
 \end{eqnarray}
 subjected to  the condition $Re \Big[\frac{K\beta \theta}{2} + \beta \sqrt{K\theta^2 \omega^2} \Big] > 0$.

The heat exchange that occurs when the system undergoes the first stage of the Stirling cycle is
\begin{equation}\label{i2}
Q_{AB}  =  U_A - U_B + \frac{1}{\beta_h} ln \Big( \frac{Z_{\omega_1,\beta_h,\theta}}{Z_{\omega_2,\beta_h,\theta}} \Big),
\end{equation}
 The partition function $Z_A$, $Z_B$ of the system can be derived using Eq.~\eqref{h1}. The internal energy $U_A$, $U_B$ is evaluated using the definition $U_i = - \partial ln Z_i/ \partial \beta_h$, where $i= A, \, B$.

In the second phase of the cycle heat is unleashed from the system. So, the heat exchange throughout this process can be expressed as
\begin{equation}\label{i4}
Q_{BC} = U_C -U_B.
\end{equation}

The third phase is a isothermal process where heat gets rejected from the system. So, the heat exchange is represented as
\begin{equation}\label{i5}
Q_{DC}  =  U_D - U_C + \frac{1}{\beta_c} ln \Big( \frac{Z_{\omega_1,\beta_c,\theta}}{Z_{\omega_2,\beta_c,\theta}} \Big),
\end{equation}

where $U_i = - \partial ln Z_i/ \partial \beta_h$ with $i= C, \, D$.

And in the last stage of the cycle the system undergoes an isochoric heat addition process. So, the heat addition to the system can be expressed as 
\begin{equation}\label{i6}
Q_{DA}  = U_A - U_D.
\end{equation}

The work done for the cycle is described as $W_{tot_{NC}} = Q_{AB} + Q_{BC} + Q_{CD} + Q_{DA}$. The efficiency of the Stirling heat cycle  for the coupled harmonic oscillator as the working substance in  NC phase space can be derived from Eq.~\eqref{i2}, ~\eqref{i4}, ~\eqref{i5} and ~\eqref{i6}. It is expressed as 
\begin{eqnarray} \label{i7} \nonumber
\eta_{Stir_{NC}}= 1 + \frac{Q_{BC} + Q_{CD}}{Q_{DA} + Q_{AB}}.
\end{eqnarray}

\begin{figure}[h]
\center
  \includegraphics[width=1.0\columnwidth]{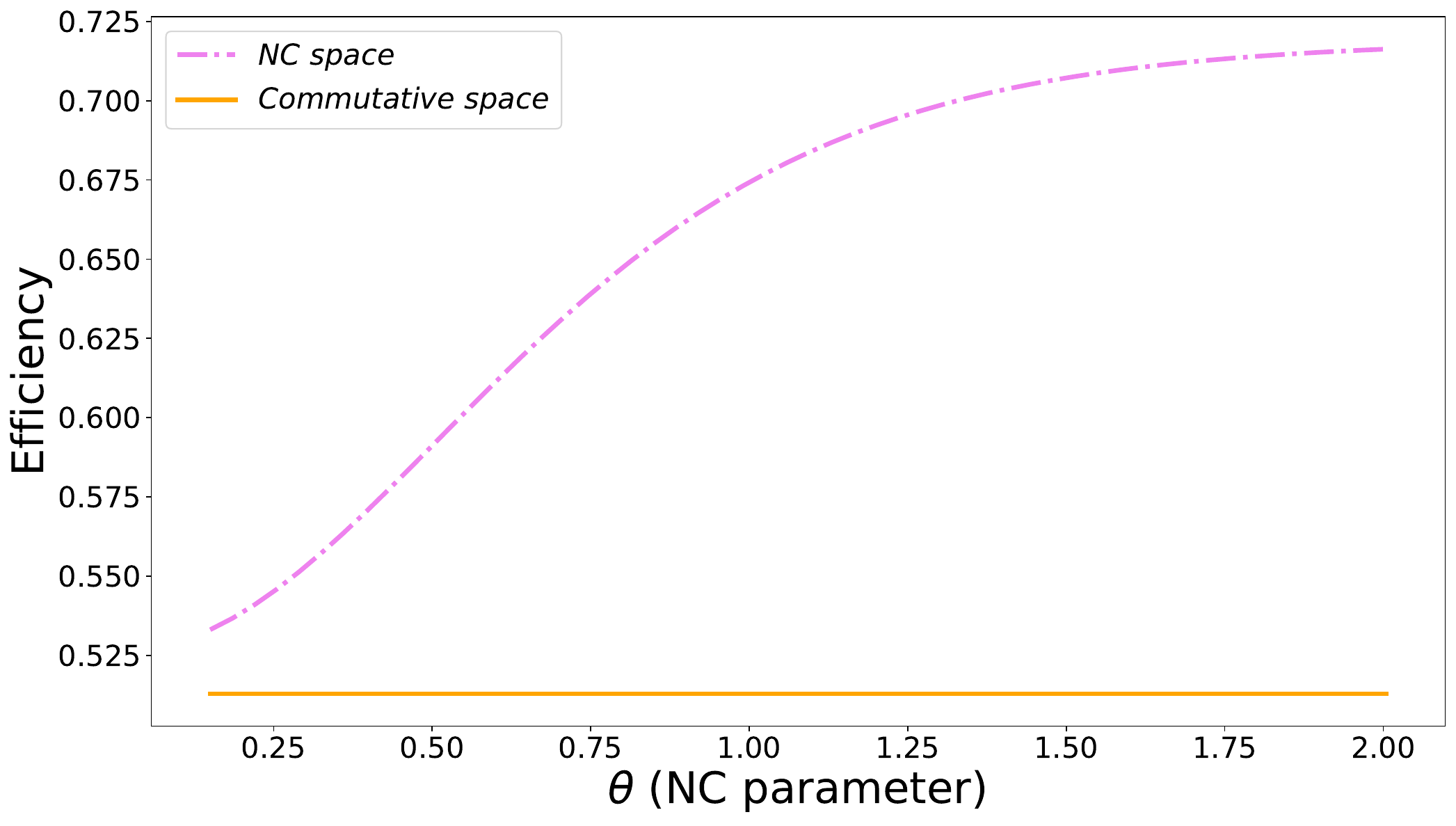}
  \caption{(Color online) Efficiency of the Stirling cycle with coupled HO as the working  substance For a constant coupling factor is shown. The violet dash-dot curve depicts the change of the efficiency of the engine with respect to the NC parameter for a constant coupling factor $\zeta= 2$. The orange solid line describes the efficiency of the commutative space with the constant coupling constant $\zeta=2$.}
  \label{fig7}
  \end{figure}

Similar to the commutative space model the hot reservoir temperature is $T_h = 4K$ and that of the cold reservoir temperature $T_c = 1K$. We take the frequency for the coupled oscillator as $\omega_1= 4$ \textcolor{black}{MHZ} and $\omega_2= 2$ \textcolor{black}{MHZ} for the evaluation of the efficiency of the Stirling cycle with respect to the variation of the NC space parameter. For our analysis, we fix the coupling strength to $\zeta = 2$ and the constant at $K= 0.25$ with NC parameter bound by \textcolor{black}{$\theta \leq 2\times 10^{-40} m^2$ }. The efficiency of the Stirling cycle (as shown in Fig.~(\ref{fig7}) in the NC phase space with coupled HO as the working medium increases with the increase of the NC parameter and attains a steady state after a certain value of the NC parameter. The efficiency of the engine is near to the efficiency of the ideal engine cycle. So, we can infer that the NC phase space provides a boost on the efficiency of the engine. In Fig.~\ref{fig7} we have shown the efficiency of the Stirling cycle in the commutative space (depicted by the solid line in the graph) for the coupling strength $\zeta=2$. The variation of the efficiency with the NC parameter for the same coupling parameter is depicted by the dash-dot curve in the graph. The efficiency of the engine boils down to the efficiency of the commutative space when the NC parameter $\theta \rightarrow 0$. So, this satisfies the condition that the results of the  system  in NC space should reduce to  the commutative space when the NC parameter is close to zero. We can infer from the graph that the engine in the NC space gets a boost for the NC parameter. Even the comparison of the efficiency of the Otto cycle with that of the Stirling cycle conveys that the working model provides a more boost on the efficiency of the Stirling cycle than that of the Otto cycle.

%%%%%%%%%%%%%%%%%%%%%%%%%%%%%%%%%%%%%%%%%%%%%%%%%%%%%%%%%%%%%%%%%%%%%%%%%%%%

\subsection{Generalized non-commutative phase space}\label{sec4c}

In the case of non-commutative phase space, the coupled oscillators as the working substances give a boost to the efficiency with respect to the NC space parameter of the system. The efficiency increases with the increase of the NC parameter and shows a steady efficiency for the higher values. Now, we will study how the generalized NC phase space affects the thermodynamic process. The Hamiltonian of the two coupled harmonic oscillators in the NC phase space is described in Eq.~\eqref{c2}. The energy eigenvalues are described in Eq.~\eqref{c5}. Following the same methodology, as used in the case of non-commutative space we will analyze the quantum Stirling cycle in the NC phase space.

In generalized NC phase space, the Hamiltonian is separated into two part as shown in Eq.~\eqref{c2}. During the first stage of the cycle, i.e, the isothermal process the individual Hamiltonian changes from the initial values from $H^{(1)}$ to $H^{(2)}$ to keep the system in thermal equilibrium with the hot reservoir.  The total Hamiltonian of the system is the sum of the effect of these two Hamiltonian. The partition function for this system when evaluated results to
\begin{equation}\label{j1}
 Z = -\frac{2 e^{\zeta-\frac{1}{2}\beta \omega \cosh(\zeta) [4 + 2K(-K \gamma + \xi)^2 sech(\zeta)]^{1/2}}}{\omega^2 \beta^2[-2e^{\zeta} + 2e^{2\zeta}K^2\gamma \xi - K (K^2 \gamma^2 +\xi^2)]},
 \end{equation}
 subjected to the condition $Re \Big[ e^{\zeta} \beta \omega \Big( \sqrt{1+ \frac{e^{\zeta} K (-K\gamma + \xi)^2}{1+ e^{2\zeta}}} + e^{2\zeta} \sqrt{1+ \frac{e^{\zeta} K (-K\gamma + \xi)^2}{1+ e^{2\zeta}}} - \sqrt{1+ \frac{e^{\zeta} K (-K\gamma + \xi)^2}{-1+ e^{2\zeta}}} + e^{2\zeta} \sqrt{1+ \frac{e^{\zeta} K (-K\gamma + \xi)^2}{-1+ e^{2\zeta}}} \Big) > 0 \Big]$

The heat exchange that takes place when the system undergoes the first stage of the Stirling cycle is
\begin{equation}\label{j2}
Q_{AB}  =  U_A - U_B + \frac{1}{\beta_h} ln \Big( \frac{Z_{\omega_1,\beta_h,\gamma,\xi}}{Z_{\omega_2,\beta_h,\gamma,\xi}} \Big),
\end{equation}
 The partition function $Z_A$, $Z_B$ of the system can be assessed using Eq.~\eqref{h1}. The internal energy $U_A$, $U_B$ is developed using the definition $U_i = - \partial ln Z_i/ \partial \beta_h$, where $i= A, \, B$.

The Hamiltonian remains  at $H^{(2)}$ while  the  temperature  of  the  system  decreases from $T_h$ to $T_c$ during the second phase of the cycle. As a result, heat is removed by the system to the reservoir and it can be mathematically defined as
\begin{equation}\label{j3}
Q_{BC} = U_C -U_B.
\end{equation}

In the third stage, the system remains coupled to the hot reservoir at temperature $T_c$ and the quasi-static changes in the Hamiltonian is depicted by the change of the Hamiltonian from $H^{(2)}$ to $H^{(1)}$.  Heat exchange for this phase of the cycle is given as
\begin{equation}\label{j4}
Q_{DC}  =  U_D - U_C + \frac{1}{\beta_c} ln \Big( \frac{Z_{\omega_1,\beta_c,\gamma,\xi}}{Z_{\omega_2,\beta_c,\gamma,\xi}} \Big),
\end{equation}
where $U_i = - \partial ln Z_i/ \partial \beta_h$ with $i= C, \, D$.

\begin{figure}[h]
\center
  \includegraphics[width=1.0\columnwidth]{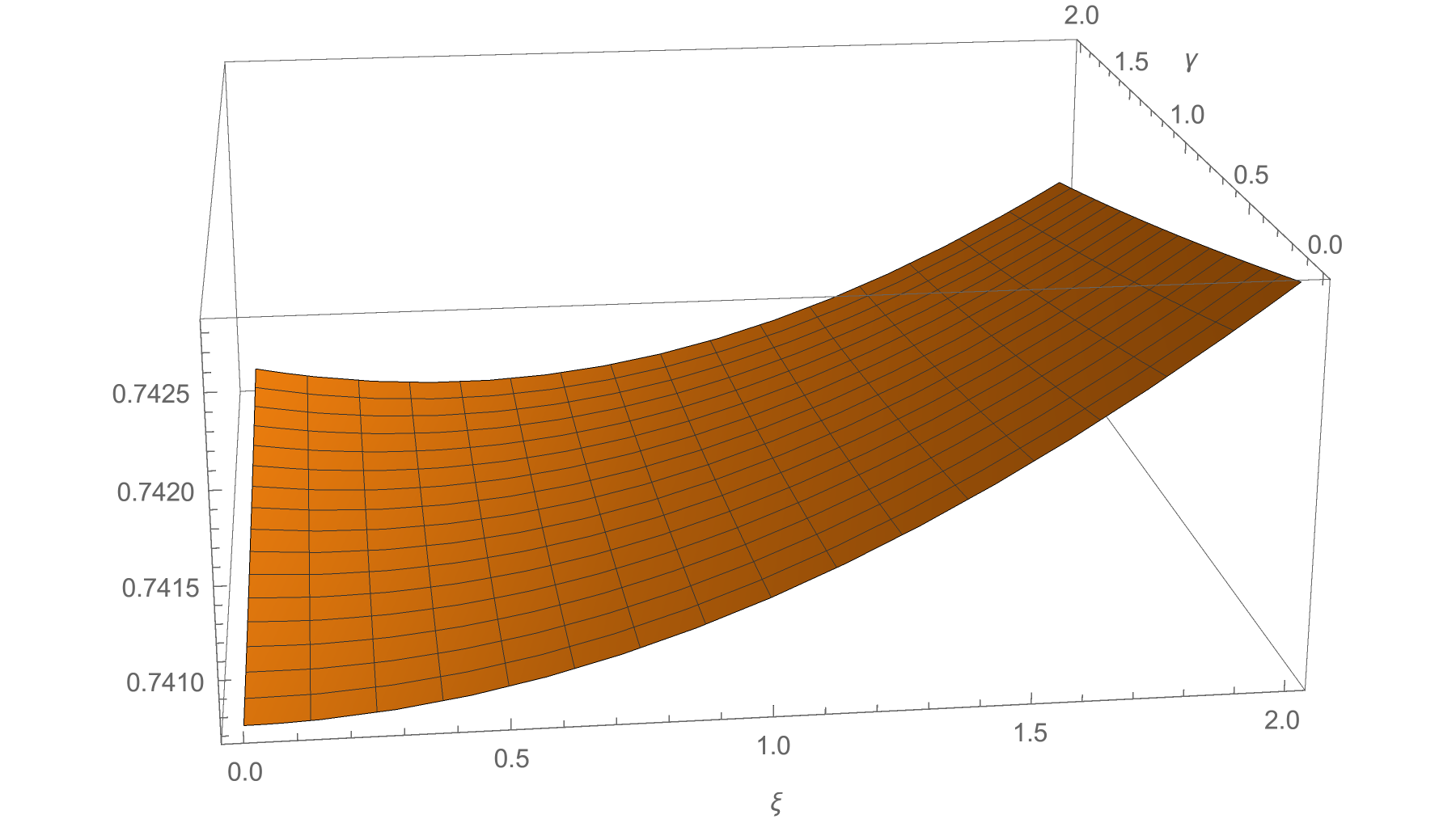}
  \caption{(Color online) Efficiency of the Stirling cycle with coupled HO as the working  substance for generalized non-commutative phase space.}
  \label{fig8}
  \end{figure}

During the fourth stage of the cycle in generalized NC space, the  system  Hamiltonian stays as it is in $H^{(1)}$ while the  temperature  changes  from $T_c$ to $T_h$ as the system is reverted back to the initial stage of the cycle. The heat exchange throughout this stage of this cycle is described as 
\begin{equation}\label{j5}
Q_{DA}  = U_A - U_D.
\end{equation}

The grand total work done for the cycle is $W_{tot_{GNC}} = Q_{AB} + Q_{BC} + Q_{CD} + Q_{DA}$. The efficiency of the Stirling heat cycle  for the coupled harmonic oscillator as the working substance in generalized  NC phase space can be derived from Eq.~\eqref{j2}, ~\eqref{j3}, ~\eqref{j4} and ~\eqref{j5}. It is expressed as 
\begin{eqnarray} \label{j6} \nonumber
\eta_{Stir_{GNC}}= 1 + \frac{Q_{BC} + Q_{CD}}{Q_{DA} + Q_{AB}}.
\end{eqnarray}

Following the same values of the parameters as done in the case of non-commutative space we considered the hot reservoir temperature $T_h = 4K$ and that of the cold reservoir temperature $T_c = 1K$ for the analysis of the Stirling cycle in generalized NC phase space. The frequency for the coupled oscillator is taken as $\omega_1= 4$ \textcolor{black}{MHz} and $\omega_2= 2$ \textcolor{black}{MHz} for the evaluation of the efficiency of the Stirling cycle with respect to the variation of the generalized NC space parameters. Throughout the process the coupling strength is $\zeta = 2$ and the constant is $K= 0.25$. The NC parameter \textcolor{black}{$\gamma \leq 2\times 10^{-40}m^2$ } and \textcolor{black}{$\xi \leq 2 \times 10^{-61}m^2/s^2$ } are bounded within this values. The three-dimensional plot (in Fig.~(\ref{fig8})) shows the variation of the efficiency of the Stirling cycle with the coupled harmonic oscillator as the working substance in the generalized NC phase space. The variation in the parameter $\xi$ and $\gamma$ in the graph shows a prominent effect on the efficiency of the engine in this space structure model. The efficiency of the stirling cycle for $\zeta = 2$ in commutative space is about $0.63$. With the same value of the coupling constant, we can see that the efficiency of the engine is high in the generalized NC space model. This indicates that the deformation of the space, which is depicted by the non-commutative parameter of both the coordinate and the momentum space influences the efficiency. If we compare the efficiency of the Stirling cycle with that of the Otto cycle we visualize that the maximum attainable efficiency for both the cycle is near about the same, but a slightly higher for the Stirling cycle.

\textcolor{black}{As described in the recent work~\cite{sdey}, the $\theta$ parameter of the NC space system is currently bounded by $\theta \sim 10^{-8} GeV^{-2}$ where they have used a quantum optical model of the present technology for the analysis. For our system, the parameter is within the range $2\times 10^{-8} GeV^{-2}$. So, optomechanical or optical system can be used for the investigation of the NC space effect on the heat engine.}

%%%%%%%%%%%%%%%%%%%%%%%%%%%%%%%%%%%%%%%%%%%%%%%%%%%%%%%%%%%%%%%%%%%%%%%%%%%%%

\section{Discussion and conclusion}\label{sec5}

To conclude, we analyzed quantum heat engines with coupled harmonic oscillators as the working medium for the commutative and the non-commutative space. The coupled harmonic oscillator in non-commutative phase space out-performs the oscillator in the commutative space in terms of the efficiency of both the quantum cycles that are being analyzed. In the case of the Otto cycle, the efficiency \textcolor{black}{increases with the increase of the NC parameter and as the NC parameter tends to zero the efficiency is equivalent to the efficiency of the commutative space}. Whereas in the case of the Stirling cycle we encounter a steep boost with the increase of the NC parameter. It tends to reach the efficiency of the ideal cycle. So, we can infer that the working medium considered for our analysis is an effectual working substance for the Stirling cycle than that of the quantum Otto cycle. Even in the case of generalized NC phase space, the efficiency gets a catalytic effect (boost) for the NC parameter over the commutative phase space. The non-linearity that we encounter in the Hamiltonian is the consequence of the NC parameter of the non-commutative phase space. 

\textcolor{black}{Analysis of NC QFT using the present quantum technology is in its amateur stage. Some of the works~\cite{ipik,sdey,mkho} have proposed some working models with quantum optics and optomechanical systems for the analysis of the presence of NC space. So, one can design the resonators (as described in the work~\cite{rv12}) with the annihilation and the creation operators of the high frequency and low frequency in the NC space. The resonators for the non-commutative space can be designed with the help of the proposed concept in the work~\cite{mkho}. The change in the resonator will result in a modified Hamiltonian of the system, which will have the NC space parameter as a variable. So, we can expect that NC space can provide an improvement in the efficiency of the heat engine. } 

Some of the previous works~\cite{san1,chatt123} have explored the influence of NC space in thermodynamic cycles. They have shown that the NC space provides a catalytic effect (boost) to the efficiency of the engine as well as the refrigerator models. Most of the analysis have claimed that the system is model dependent. So to provide a generic statement that the high efficiency of engine in NC space is universal, further analysis in this direction are required.

For the analysis and implementation of the quantum Otto cycle, one has to maintain a quasi-static adiabatic process to prevent generation of coherence in the Hamiltonian of the considered system. This should be kept in mind while the execution of the adiabatic process, so that one can prevent the mean population change. During the thermalization process of the cycle, to achieve the thermal equilibrium state where the system is coupled with the reservoirs, the system must stay coupled with the bath for a longer period of time. It will be interesting to analyze various thermodynamic processes by using the general form of coupling of the harmonic oscillator as the working substance.

The NC phase space can be an effectual resource for different application areas of the quantum theory~\cite{mhu}, which needs further exploration. Various other coupled working medium is used for the analysis of quantum cycles~\cite{tzh}. One can make use of the NC phase space structure to analyze these models for the cycles and even can be extended for exploring non-Markovian reservoirs. For our analysis, we have focused on quantum heat cycles. One can study the effect of the non-commutative phase space on the coefficient of performance of the quantum refrigerator cycles for coupled oscillators and even for other working substances. The analysis of the existing thermodynamic cycles in NC phase space is required to provide the generic statement about the boost it yields to the efficiency of the cycles. The analysis of the irreversible and continuous cycles~\cite{liu1,dalk,kos12,klat} and quantum phase transition in NC phase space needs exploration to visualize the effect of the NC parameter in different thermodynamic processes. The challenging task in the NC space is to analyze the NC spacetime in the relativistic regime. Recent work~\cite{todo} have analyzed the different potential problem in the NC spacetime with relativistic correction. This gives us the insight to analyze the thermodynamic process in the relativistic realm of NC phase space which needs exploration.

One can utilize generalized uncertainty principle to instigate a bound not only in the efficiency of the different cycle but even to the various thermodynamic process by getting motivated from previous works~\cite{chatt,chatt1,chatt2}. The experimental realization of the NC phase space~\cite{sdey,mkho} with our existing technology will provide a boost for the analysis of thermodynamic processes and quantum information theory in NC spacetime. Dey and Hussin~\cite{deyyy} in their work have shown that non-commutative systems result in more entanglement than the usual quantum systems. The experimental validation of this will provide a boost in the study of entanglement theory and its application to the different area of quantum information theory.

\section*{Acknowledgment}

The authors gratefully acknowledge for the useful discussions and suggestions from Dr. Raam Uzdin, Senior Lecturer (Assist. Prof.) at the Hebrew University of Jerusalem, Faculty of Science, Fritz Haber Center for Molecular Dynamics Institute of Chemistry.

%%%%%%%%%%%%%%%%%%%%%%%%%%%%%%%%%%%%%%%%%%%%%%%%%%%%%%%%%%%%%%%%%%%%%%%%

%%%%%%%%%%%%%%%%%%%%%%%%%%%%%%%%%%%%%%%%%%%%%%%%%%%%%%%%%%%%%%%%%%%%%%%%
\bibliographystyle{h-physrev4}

%%%%%%%%%%%%%%%%%%%%%%%%%%%%%%%%%%%%%%%%%%%%%%%%%%%%%%%%%%%%%%%%%

%%%%%%%%%%%%%%%%%%%%%%%%%%%%%%%%%%%%%%%%%%%%%%%%%%%%%%%%%%%%%%%%%

\end{document}